\documentclass[sn-mathphys, Numbered, iicol]{sn-jnl}
\usepackage{graphicx}%
\usepackage{multirow}%
\usepackage{amsmath,amssymb,amsfonts}%
\usepackage{amsthm}%
\usepackage{mathrsfs}%
\usepackage[title]{appendix}%
\usepackage{xcolor}%
\usepackage{textcomp}%
\usepackage{manyfoot}%
\usepackage{booktabs}%
\usepackage{algorithm}%
\usepackage{algorithmicx}%
\usepackage{algpseudocode}%
\usepackage{listings}%
\usepackage{cleveref}
\usepackage{ulem}
\usepackage{xspace}
\usepackage[version=4]{mhchem}
\usepackage{subcaption}
\usepackage{bbm}
\usepackage{siunitx}
\makeatletter

\makeatletter
\newcommand*{\myfnsymbolsingle}[1]{%
  \ensuremath{%
    \ifcase#1
    \or 
      \S%
    \or 
      \dagger
    \or 
      a
    \or 
      b
    \or 
      c
    \or 
      d
    \else 
      \@ctrerr  
    \fi
  }%
}   
\makeatother
\usepackage{alphalph}
\usepackage{fnpct} 
\usepackage{xr}

\newcommand{\corres}[1]{}

\Crefname{figure}{\text{Fig.}}{\text{Figs.}}
\Crefname{section}{\text{Section}}{\text{Sections}}
\Crefname{equation}{\text{Eq.}}{\text{Eqs.}}

\usepackage{amsmath,amsfonts,bm}

\def\eqref#1{equation~\ref{#1}}

\def\1{\bm{1}}

\DeclareMathAlphabet{\mathsfit}{\encodingdefault}{\sfdefault}{m}{sl}
\SetMathAlphabet{\mathsfit}{bold}{\encodingdefault}{\sfdefault}{bx}{n}

\begin{document}

\title[Learning Electron Density with Virtual Atoms]{Learning Electron Density with Virtual Atoms for Transferable Atomistic Representations}

\author[1]{\fnm{Yuzhi} \sur{Liu}}
\author[1,2]{\fnm{Xinzijian} \sur{Liu}}
\author[1,3]{\fnm{Chengqian} \sur{Zhang}}
\author[1]{\fnm{Anyang} \sur{Peng}}
\author[1,2,3]{\fnm{Duo} \sur{Zhang}}
\author*[4,5]{\fnm{Han} \sur{Wang}}
\email{wang\_han@iapcm.ac.cn}

\affil[1]{AI for Science Institute, Beijing 100080, P. R.~China}
\affil[2]{DP Technology, Beijing 100080, P. R.~China}
\affil[3]{Academy for Advanced Interdisciplinary Studies, Peking University, Beijing 100871, P. R.~China}
\affil[4]{National Key Laboratory of Computational Physics, Institute of Applied Physics and Computational Mathematics, Fenghao East Road 2, Beijing 100094, P. R.~China}
\affil[5]{HEDPS, CAPT, College of Engineering, Peking University, Beijing 100871, P. R.~China}



\abstract{
Electron density is a fundamental quantity of electronic structure, 
yet most machine learning density models treat it solely as a reconstruction target and are developed separately from machine learning interatomic potentials.
Here, we introduce a virtual-atom formulation within the DPA3 framework,
in which density query points are represented as virtual nodes and processed by the same parameterized encoder used for real atoms.
An asymmetric neighbor construction allows virtual atoms to receive information from nearby real atoms without altering the real-atom interaction graph.
The resulting framework supports density, energy, and force prediction through task-specific readout heads.
The proposed model achieves state-of-the-art density reconstruction accuracy
on the Materials Project benchmark, matches the best reported performance on nickel--manganese--cobalt battery cathode
materials, and is competitive with dedicated density models on QM9.
When density labels are used jointly with energy and force supervision,
the model improves in-domain accuracy, makes local-environment information more
accessible under linear probing, and reduces zero-shot force errors on larger
QM40 molecules. Density-only checkpoints also provide useful initializations
for downstream energy--force learning under the fixed fine-tuning protocol.
Together, these results support electron density as a source of electronic-
structure supervision for interatomic-potential representations.
}

\maketitle

\section{Introduction}\label{sec:Intro}

Machine learning interatomic potentials (MLIPs) have transformed atomistic simulation by enabling 
large-scale molecular dynamics and materials discovery
with near quantum mechanical 
accuracy~\citep{zhang2018deep,batzner20223,batatia2022mace,zhang2025graph,
behler2007generalized,bartok2010gaussian,schutt2017schnet,
chen2022universal,deng2023chgnet,yang2024mattersim,zhang2024dpa,liu2026mixture}.
In these models, local atomic environments are 
encoded into atomic representations and 
mapped to energies and forces.
Despite their remarkable success 
in fitting potential-energy surfaces, 
MLIPs are typically supervised by 
response quantities derived from the 
Born--Oppenheimer energy rather 
than by the electronic structure itself.
Consequently, the learned representations capture the energetics
of nuclear rearrangements but are not explicitly constrained by the underlying
electron density. Transferability across chemically diverse systems remains
challenging for current MLIPs~\citep{deng2024overcoming,focassio2024performance,kim2026optimizing},
and whether explicit electronic-structure supervision can alleviate such limitations
has received comparatively little attention.

Electron density offers a natural route to inject such electronic structure information into atomistic learning.
As the fundamental variable of density functional theory (DFT), the ground-state electron density determines
all ground-state properties of a many-electron 
system~\citep{hohenberg1964inhomogeneous,kohn1965self}, 
and, together with the nuclear interaction terms, enters the formal expression for atomic forces through the Hellmann--Feynman theorem.
Unlike energies and forces, electron density provides spatially
resolved information about the electronic structure around atoms and in interstitial regions.
It therefore offers a richer supervision signal for learning atomistic representations.
Motivated by this, machine learning models for electron density have advanced rapidly.
Grid-based methods treat density as a volumetric field and predict values at sampled query points 
conditioned on atomic 
environments~\citep{jorgensen2020deepdft,jorgensen2022equivariant,
lv2023deep,zepeda2021deep,focassio2023linear,chandrasekaran2019solving,koker2024higher}.
Complementary approaches expand the density in atom-centered basis functions and learn the corresponding
coefficients~\citep{lewis2021learning,grisafi2018transferable,fabrizio2019electron,
rackers2023recipe,fu2024recipe,klockow2026function}, in some cases with additional
basis functions centered away from the atoms~\citep{fu2024recipe,elsborg2025electra}.
More recently, hybrid formulations have
combined grid-based sampling with atom-centered
decompositions~\citep{qin2026eac}. Machine-learned densities have also been used as inputs to energy models,
from machine-learned Hohenberg--Kohn maps~\citep{brockherde2017bypassing,bogojeski2020quantum}
to molecular dynamics driven by energies predicted from machine-learned densities~\citep{bogojeski2026enhancing}.
These advances demonstrate that electron 
density is an information-rich target that can 
be predicted with high accuracy.

Although electron densities, energies, and forces are often generated within the same electronic-structure calculations,
density reconstruction and interatomic-potential learning have generally been optimized as separate tasks, aside from approaches that use a predicted density as an explicit input to an energy model.
Most density models focus on reconstructing a spatial field, whereas MLIPs learn atomic representations directly from energy and force supervision.
The information in the electron density is therefore rarely used to shape the representations employed by a potential.
High-throughput datasets such as OMol25~\citep{levine2025open} increasingly make these complementary labels available, raising the question of whether electron density can provide transferable supervision for energy and force prediction through density pretraining or multi-task EF+density training.

In this work, we address this question by reformulating electron density learning as an atomistic
representation-learning task within the DPA3 framework~\citep{zhang2025graph}.
DPA3 is a graph neural network architecture in which local atomic environments are encoded into atomic representations and then 
mapped to physical observables through task-specific readout networks.
This structure provides a natural basis for unifying density, energy, and force prediction.
Specifically, we place auxiliary nodes at prescribed density-query locations
and combine them with real atoms in a merged atomic system. These nodes play the
role of query points in grid-based density models, but are processed by the
DPA3 representation network together with the real atoms; we therefore refer
to them as virtual atoms. Virtual atoms receive information from nearby real
atoms, while real atoms do not receive messages from virtual atoms. This
asymmetry preserves the original real-atom interaction graph and allows the
same encoder parameters to support density and energy--force prediction with
separate task-specific readouts.
The one-way flow from atoms to density-query nodes follows the probe-node construction used in grid-based density models~\citep{jorgensen2020deepdft,jorgensen2022equivariant,koker2024higher}.
The distinguishing feature here is that virtual and real centers are processed by the same encoder parameters, allowing density supervision to act directly on the representation used for energy and force prediction.

This construction is physically motivated by the relationship between electron density, energy, and forces
in the Born--Oppenheimer framework~\citep{sheng2026density}.
Within density functional theory, the ground-state electron density is the first functional derivative of the 
electronic energy with respect to the external potential, whereas atomic forces are derivatives of the Born--Oppenheimer energy with respect to nuclear coordinates.
This formal relationship does not imply that density reconstruction uniquely determines atomic forces, but it suggests that density supervision may encourage the encoder to retain spatially resolved electronic structure information relevant to force learning.

We evaluate this idea at three levels across molecular and periodic systems.
First, we test whether the virtual-atom formulation can reconstruct electron density accurately.
The proposed model achieves state-of-the-art density reconstruction accuracy
on the Materials Project (MP) dataset~\citep{shen2022representation,jain2013materials},
matches the best reported performance on nickel--manganese--cobalt (NMC) battery cathode materials~\citep{jorgensen2022nmc},
and is competitive with dedicated density models on the QM9 dataset~\citep{ruddigkeit2012enumeration,ramakrishnan2014quantum,jorgensen2022qm9}.
Second, we investigate the effect of multi-task EF+density training on the learned representations, in-domain energy and force accuracy, and zero-shot transfer.
Density supervision makes local-environment information more accessible under linear probing, improves in-domain prediction, and substantially reduces zero-shot force errors on an element-matched, ORCA-relabeled subset of QM40 containing molecules larger than those encountered during training.
Third, we examine whether density-only pretraining provides a useful
initialization for downstream energy--force learning. Under a fixed fine-tuning
protocol, density-pretrained checkpoints generally yield lower validation
errors than random initialization, particularly in low-data regimes.

\section{Methodology}\label{sec:method}

\begin{figure*}
    \centering
    \includegraphics[width=0.9\linewidth]{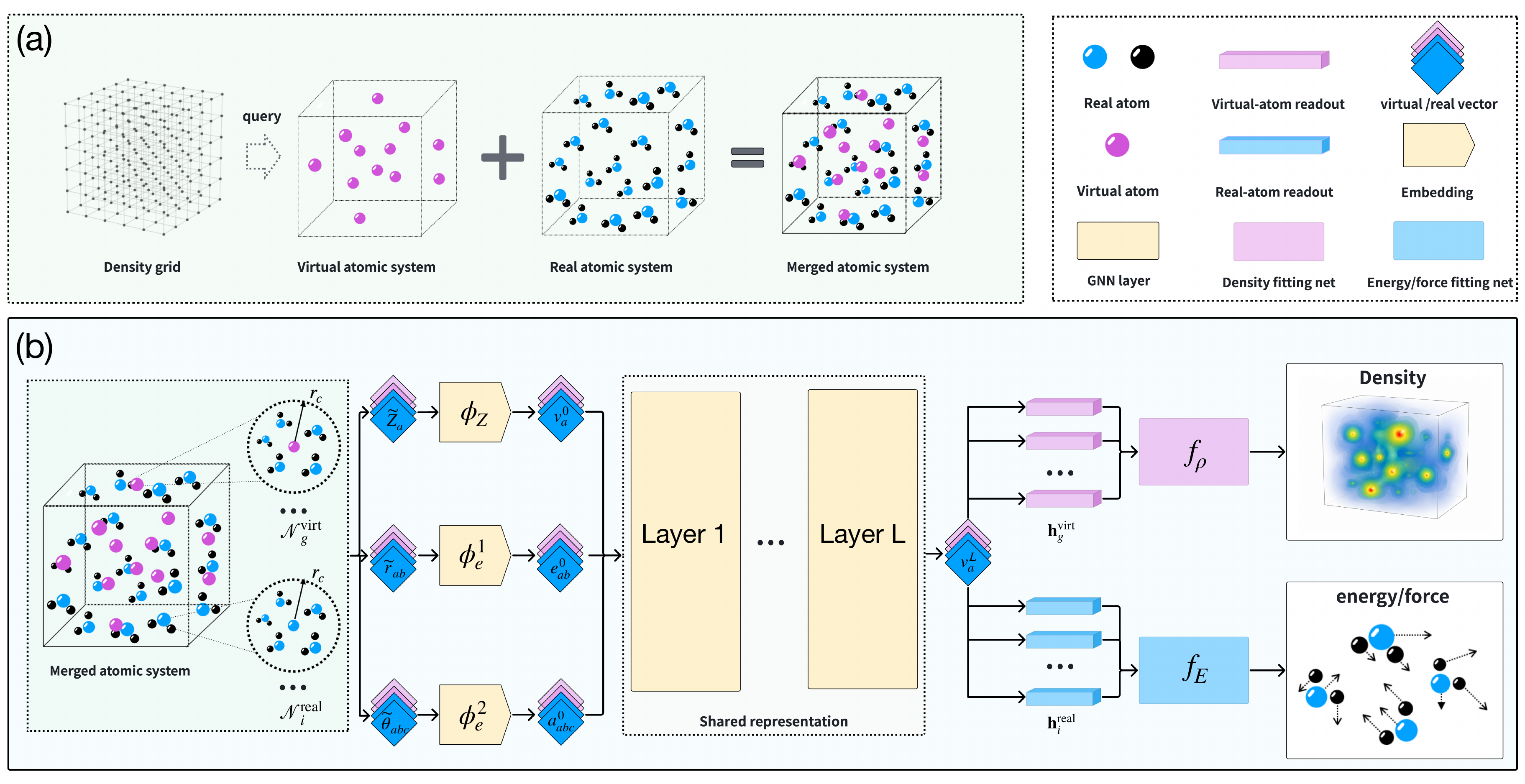}
    \caption{
Virtual-atom formulation for joint density and energy--force learning.
(a) Density query points become typed virtual atoms that receive information from nearby real atoms without altering the real-atom interaction graph.
(b) Virtual and real centers share DPA3 encoder parameters and use task-specific readouts for density and energy prediction; forces are obtained as energy derivatives.
    }
    \label{fig:figure1}
\end{figure*}

\subsection{Virtual-Atom Graph}

Figure~\ref{fig:figure1}(a) illustrates the construction of the merged atomic system.
For each atomic configuration, let $\rho(\mathbf r)$ denote the reference electron density at
position $\mathbf r\in\mathbb R^3$.
We evaluate the density at a finite set of spatial query points, or density-grid points,
\begin{equation}
\mathcal{Q}
=
\{\mathbf r_g\in\mathbb R^3\}_{g=1}^{N_g}.
\end{equation}
Here, $g$ indexes the query points, $\mathbf r_g$ is the Cartesian position of the
$g$-th query point, and $N_g=|\mathcal Q|$ is the total number of query points.
Instead of treating these query points as external field locations, we reinterpret each query point as an atom-like node.
Each query point $\mathbf r_g$ is assigned a dedicated virtual atom type $Z_{\mathrm v}$, forming a virtual atomic system
\begin{equation}
\mathcal{V}
=
\{(Z_{\mathrm v},\mathbf r_g)\}_{g=1}^{N_g}.
\end{equation}
Each virtual atom is therefore 
a density probe: its supervised label is the reference 
density value $\rho(\mathbf r_g)$ evaluated at 
the same spatial position.

Given a real atomic system
\begin{equation}
\mathcal{A}
=
\{(Z_i,\mathbf R_i)\}_{i=1}^{N},
\end{equation}
where $i$ indexes the real atoms, $Z_i$ is the chemical species of atom $i$,
$\mathbf R_i\in\mathbb R^3$ is its Cartesian position, and $N$ is the number of real atoms,
the virtual and real atomic systems are 
combined into a merged atomic system,
\begin{equation}
\widetilde{\mathcal A}
=
\mathcal V
\cup
\mathcal A.
\end{equation}
Using a virtual-first ordering, the corresponding ordered coordinate and atom-type arrays are
\begin{align}
\widetilde{\mathbf R}
&=
(\mathbf r_1,\ldots,\mathbf r_{N_g},
\mathbf R_1,\ldots,\mathbf R_N),
\\
\widetilde{\mathbf Z}
&=
(Z_{\mathrm v},\ldots,Z_{\mathrm v},
Z_1,\ldots,Z_N).
\end{align}

The virtual atoms are introduced only as density 
probes and should not modify the physical 
interaction topology among real atoms.
We therefore use an asymmetric neighbor construction.
Let $r_c$ denote the neighbor cutoff radius and $\|\cdot\|_2$ the Euclidean norm.
For each real atom $i$, 
the neighbor environment contains only real atoms,
\begin{equation}
\mathcal N_i^{\mathrm{real}}
=
\{
j\in\{1,\ldots,N\}\setminus\{i\}
\mid
\|\mathbf R_j-\mathbf R_i\|_2<r_c
\}.
\end{equation}
For each virtual atom $g$, 
the neighbor environment contains nearby real atoms,
\begin{equation}
\mathcal N_g^{\mathrm{virt}}
=
\{
j\in\{1,\ldots,N\}
\mid
\|\mathbf R_j-\mathbf r_g\|_2<r_c
\}.
\end{equation}
Collecting the neighbor environments in the same virtual-first ordering as
$\widetilde{\mathbf R}$ and $\widetilde{\mathbf Z}$, we define
\begin{equation}
\widetilde{\mathcal N}
=
(\mathcal N_1^{\mathrm{virt}},\ldots,\mathcal N_{N_g}^{\mathrm{virt}},
\mathcal N_1^{\mathrm{real}},\ldots,\mathcal N_N^{\mathrm{real}}),
\end{equation}
Here $p\in\{1,\ldots,N_g+N\}$ indexes nodes in the merged atomic system:
$p=g$ corresponds to virtual node $g$ for $1\leq p\leq N_g$, whereas
$p=N_g+i$ corresponds to real atom $i$ for $N_g<p\leq N_g+N$.
Thus, $\widetilde{\mathcal N}_p$ is the neighbor environment of the $p$-th node.
No virtual--virtual edges are introduced, and virtual atoms are never included as neighbors of real atoms.
Consequently, virtual atoms receive information from their surrounding real atoms, whereas 
the real-atom energy and force pathway remains identical to that of the original DPA3 potential.

This asymmetric information flow follows the probe-node construction of grid-based density models~\citep{jorgensen2020deepdft,jorgensen2022equivariant,koker2024higher}, in which density-query points receive information from neighboring atoms.
Virtual atoms in the present work differ from the virtual or floating orbitals used by basis-function density models~\citep{fu2024recipe,elsborg2025electra}: those orbitals are additional density-expansion centers, whereas the virtual atoms here are query locations at which the density is evaluated.
Our construction further processes virtual and real centers with the same encoder parameters, distinguished by the dedicated atom type $Z_{\mathrm v}$.

A virtual atom carries no nuclear charge, mass, or energy contribution; it serves only as a typed query node at which the electron density is evaluated.
In this merged atomic system, density query locations become atom-like centers processed by the same parameterized DPA3 encoder as real atoms.
The encoder maps each center and its neighbor environment to a 
latent representation; the density label remains attached to the original 
spatial position $\mathbf r_g$.
The asymmetric construction therefore creates two types of centers within the same parameterized encoder: virtual centers used to query the density and real-atom centers used to construct the potential.
Density learning can update the shared encoder parameters without introducing virtual-atom interactions into the physical energy and force pathway.

\subsection{Shared Encoder and Task-Specific Readouts}

The merged atomic system is encoded using the DPA3 representation, as illustrated in Fig.~\ref{fig:figure1}(b).
We write the input graph as
\begin{equation}
\widetilde{\mathcal G}
=
(
\widetilde{\mathbf R},
\widetilde{\mathbf Z},
\widetilde{\mathcal N}
),
\end{equation}
where $\widetilde{\mathcal N}$ is the ordered family of asymmetric
neighbor environments defined above.

We index nodes in the merged atomic system by $p,q,s$.
  For each center node $p$, DPA3 constructs three types of
  geometric inputs from its neighbor environment $\widetilde{\mathcal N}_p$: node, edge, and angular inputs.
  The node input is the atom type $\widetilde Z_p$, where virtual atoms use the dedicated type $Z_{\mathrm v}$.
  For a neighbor $q\in\widetilde{\mathcal N}_p$, the edge input is the intercenter distance
  $\widetilde r_{pq}=\|\widetilde{\mathbf R}_q-\widetilde{\mathbf R}_p\|_2$.
  For a neighbor pair $q,s\in\widetilde{\mathcal N}_p$, the angular input
  $\widetilde\theta_{pqs}$ is the angle between
  $\widetilde{\mathbf R}_q-\widetilde{\mathbf R}_p$ and
  $\widetilde{\mathbf R}_s-\widetilde{\mathbf R}_p$.
  For real centers, these node, edge, and angular inputs 
  are constructed only from real-atom neighbors.
  For virtual centers, the corresponding edge and angular inputs are constructed 
  from the real atoms surrounding the density query point.
  Thus, virtual atoms encode the local electronic environment 
  around query locations without contributing to the real-atom interaction graph.

Following the original DPA3 formulation, these inputs are embedded as
\begin{equation}
  \begin{aligned}
  \mathbf v_p^{(0)}
  &= \phi_Z(\widetilde Z_p),
  \\
  \mathbf e_{pq}^{(0)}
  &= \phi^{1}(\widetilde r_{pq}),
  \\
  \mathbf a_{pqs}^{(0)}
  &= \phi^{2}(\widetilde\theta_{pqs}).
  \end{aligned}
\end{equation}
Here, $\phi_Z$, $\phi^1$, and $\phi^2$ are learnable embedding maps for the
atom-type, radial, and angular inputs, respectively, and
$\mathbf v_p^{(0)}$, $\mathbf e_{pq}^{(0)}$, and $\mathbf a_{pqs}^{(0)}$
are the resulting initial node, edge, and angular feature vectors.
Let $\mathbf v^{(\ell)}$, $\mathbf e^{(\ell)}$, and $\mathbf a^{(\ell)}$
denote the complete collections of node, edge, and angular features after layer $\ell$.
These features are updated through $L$ DPA3 interaction layers, for $\ell=1,\ldots,L$,
\begin{equation}
(
\mathbf v^{(\ell)},
\mathbf e^{(\ell)},
\mathbf a^{(\ell)}
)
=
\Psi^{(\ell)}
(
\mathbf v^{(\ell-1)},
\mathbf e^{(\ell-1)},
\mathbf a^{(\ell-1)}
),
\end{equation}
where $\Psi^{(\ell)}$ is the update operator of the $\ell$-th layer.
After $L$ interaction layers, the final-layer output contains the node representations,
\begin{equation}
\widetilde{\mathbf h}
=
\mathbf v^{(L)}
=
\Phi_{\mathrm{enc}}(\widetilde{\mathcal G}),
\end{equation}
where $\Phi_{\mathrm{enc}}$ denotes the shared DPA3 encoder, with its
learnable parameters suppressed for brevity, and $\widetilde{\mathbf h}$ is the
ordered collection of final node representations used by the task-specific readouts.

We separate these final node representations into virtual and real parts,
\begin{equation}
\widetilde{\mathbf h}
=
\bigl(
\mathbf h_1^{\rm virt},\ldots,\mathbf h_{N_g}^{\rm virt},
\mathbf h_1^{\rm real},\ldots,\mathbf h_N^{\rm real}
\bigr).
\end{equation}
Let $f_\rho$ and $f_E$ denote learnable scalar readout functions for density
and atomic energy, respectively; hats indicate model predictions.
Virtual-node representations are mapped to electron densities,
\begin{equation}
\hat{\rho}(\mathbf r_g)
=
f_\rho(\mathbf h_g^{\rm virt}),
\end{equation}
whereas real-atom representations are mapped to atomic energy contributions,
\begin{equation}
\hat E
=
\sum_{i=1}^{N}
f_E(\mathbf h_i^{\rm real}).
\end{equation}
Atomic forces are obtained by differentiating the predicted energy with respect to real atomic coordinates,
\begin{equation}
\hat{\mathbf F}_i
=
-
\frac{\partial \hat E}
{\partial \mathbf R_i}.
\end{equation}
The density and energy--force branches therefore operate on different node representations but share the same DPA3 encoder parameters.
Gradients from density reconstruction can consequently shape the encoder that is later used, or simultaneously optimized, for energy and force prediction.

\subsection{Density--Force Motivation}
\label{sec:density_force_signal}

The use of electron density as a supervision signal is motivated by its 
first-order relationship with atomic forces in the Born--Oppenheimer framework.
Let $\mathbf R=(\mathbf R_1,\ldots,\mathbf R_N)$ denote the complete nuclear configuration.
Following Ref.~\citep{sheng2026density}, its Born--Oppenheimer energy surface can be expressed as
\begin{equation}
E_{\mathrm{BO}}(\mathbf R)
=
E[v_{\mathbf R}]
+
V_{\mathrm{nn}}(\mathbf R),
\end{equation}
where $E[v]$ is the ground-state electronic energy as a functional of the external
potential $v$, $v_{\mathbf R}(\mathbf r)$ is the external potential generated by the
nuclei, and $V_{\mathrm{nn}}(\mathbf R)$ is the nuclear--nuclear repulsion energy.
Within density functional theory, the ground-state electron density is the first functional derivative of the electronic energy with respect to the external potential,
\begin{equation}
\rho_{\mathbf R}(\mathbf r)
=
\left.
\frac{\delta E[v]}{\delta v(\mathbf r)}
\right|_{v=v_{\mathbf R}} .
\end{equation}
The force on real atom $i$ is also a first-order derivative of the Born--Oppenheimer energy, but with respect to its nuclear coordinate,
\begin{equation}
\mathbf F_i(\mathbf R)
=
-
\frac{\partial E_{\mathrm{BO}}(\mathbf R)}
{\partial \mathbf R_i}.
\end{equation}
Applying the chain rule through the nuclear-to-potential map $\mathbf R\mapsto v_{\mathbf R}$ gives
\begin{equation}
\mathbf F_i(\mathbf R)
=
-
\int
\rho_{\mathbf R}(\mathbf r)
\frac{\partial v_{\mathbf R}(\mathbf r)}
{\partial \mathbf R_i}
\,d\mathbf r
-
\frac{\partial V_{\mathrm{nn}}(\mathbf R)}
{\partial \mathbf R_i}.
\label{eq:force_density_relation}
\end{equation}
Equation~\ref{eq:force_density_relation} 
shows that the electron density and atomic forces 
are two different responses of the same Born--Oppenheimer energy: 
the density describes how the energy responds to a change in the external potential,
whereas the force describes how it responds to a displacement of the nuclei. 
This shared dependence on the underlying electronic structure suggests 
that density supervision may encourage the model to learn features that are 
also useful for predicting forces. However, this argument does not imply 
a one-to-one correspondence between density accuracy and force accuracy, 
nor does an accurate density reconstruction by itself guarantee accurate forces. 
The relation is exact only when the density and forces refer to 
the same system and the same Hamiltonian; in our setting, 
it therefore serves as a physical motivation for joint 
learning rather than a direct quantitative relation between the two prediction errors.

\subsection{Training Objectives}

The framework supports four related training settings: density-only training, 
standard energy--force (EF) training, multi-task EF+density training, and density pretraining followed by EF fine-tuning.

For density-only training, the model is optimized using the mean absolute error (MAE)
between the predicted and reference densities at virtual-atom query locations,
\begin{equation}
\mathcal{L}_{\rho}
=
\frac{1}{N_g}
\sum_{g=1}^{N_g}
\left|
\hat{\rho}(\mathbf r_g)
-
\rho(\mathbf r_g)
\right|.
\end{equation}
This objective is used both to evaluate whether the virtual-atom formulation can 
reconstruct electron density and to obtain density-pretrained representations.
In the results, density reconstruction is reported using the normalized density MAE, 
whereas the training objective uses the unnormalized MAE above.

For standard energy--force training, only the real-atom branch is used.
The energy and force losses are
\begin{equation}
\mathcal{L}_{E}
=
\frac{1}{M}
\sum_{m=1}^{M}
\left(
\hat{E}^{(m)}
-
E^{(m)}
\right)^2,
\end{equation}
and
\begin{equation}
\mathcal{L}_{F}
=
\frac{1}{3\sum_{m=1}^{M}N_m}
\sum_{m=1}^{M}
\sum_{i=1}^{N_m}
\left\|
\hat{\mathbf F}_{i}^{(m)}
-
\mathbf F_{i}^{(m)}
\right\|^2,
\end{equation}
where $M$ is the number of configurations in a batch, $N_m$ is the number of atoms
in configuration $m$, and $i$ indexes its atoms. $E^{(m)}$ and
$\mathbf F_i^{(m)}$ are reference labels, while their hatted counterparts are predictions.
The total energy--force objective is
\begin{equation}
\mathcal{L}_{EF}
=
\lambda_E \mathcal{L}_E
+
\lambda_F \mathcal{L}_F .
\end{equation}
Here, $\lambda_E$ and $\lambda_F$ are nonnegative weights for the energy and force losses.

For multi-task EF+density training, the density and energy--force branches are optimized together,
\begin{equation}
\mathcal{L}_{\mathrm{multi}}
=
\lambda_{\rho}\mathcal{L}_{\rho}
+
\lambda_E \mathcal{L}_E
+
\lambda_F \mathcal{L}_F .
\end{equation}
Here, $\lambda_\rho$ is the nonnegative density-loss weight.
In the reported calculations, $\lambda_E$ varies from $0.2$ to $20$, $\lambda_F$ varies from $100$ to $60$, and $\lambda_\rho=1$ throughout (Supplementary Tables~2 and~3).
The density, energy, and force readouts are task-specific, but all losses update the shared DPA3 encoder parameters.
The density objective can therefore influence the real-atom representations used by the energy and force heads.

When density, energy, and force labels are available for the same configurations, all three losses are evaluated on the corresponding labeled batch.
When density labels and energy--force labels come from related but not identical datasets, we use mixed batches 
from the two sources and apply each loss only to samples for which the corresponding labels are
available.
Each standard EF training step contains 256 energy--force structures, whereas each EF+density step contains the same 256 energy--force structures together with 256 density structures.
Both models are optimized for $2\times10^6$ steps and therefore receive the same total energy--force sample exposure, $5.12\times10^8$ structure presentations.
The additional samples in the EF+density batch carry only density labels and do not increase the energy--force label budget.
In both cases, the same encoder parameters are shared across tasks, so density supervision shapes the encoder used for energy--force prediction.

\subsection{Density Pretraining and Fine-Tuning}

Beyond EF+density training, the same framework supports density pretraining followed by downstream energy--force fine-tuning.
In the pretraining stage, the DPA3 encoder and density readout are optimized using $\mathcal L_\rho$ on an electron-density dataset.
This stage trains the shared encoder parameters to extract local electronic structure around density-query locations from their neighboring atoms.

In the fine-tuning stage, the pretrained DPA3 encoder is transferred to a downstream potential learning task.
The density readout is discarded, and an energy readout is attached to the real-atom representations.
The model is then optimized using only downstream energy and force labels through $\mathcal L_{EF}$.
Unless otherwise specified, all encoder and readout parameters are updated during fine-tuning.

This procedure differs from EF+density training in that density labels are absent during downstream optimization.
Density therefore acts as a pretraining task rather than as a simultaneous auxiliary loss.
If density learning produces a transferable electronic structure representation, the pretrained
representation should provide a favorable initialization for subsequent energy and force learning, including when downstream labels are limited.

\subsection{Linear Probing}

To assess the local-environment information encoded in the learned representations, we freeze each DPA3 encoder and train a separate logistic-regression classifier on the resulting per-atom descriptors.
The probes are trained on the same subset of QM9x training atoms, constructed by sampling up to 2,000 atoms per local-environment class, and the descriptors are standardized using the probe-training data.
Local environments are grouped into ten classes according to atomic species and covalent coordination: C with two, three, or four bonded neighbors; N with one, two, or three bonded neighbors; O with one or two bonded neighbors; H; and F.
Covalent coordination is determined from interatomic distances and covalent radii, with atom pairs considered bonded when their separation is less than $1.15$ times the sum of their covalent radii.
We report balanced accuracy on the sampled QM9x training atoms and evaluate the trained probes, without further optimization, on the QM9x validation split and the ORCA-relabeled QM40 subset.

\section{Results}\label{sec:results}

We evaluate the proposed virtual-atom formulation through three complementary questions across molecular and periodic systems.
First, can the formulation reconstruct spatially resolved electron density accurately?
Second, does retaining density supervision during energy--force training improve the learned representations, in-domain prediction, and zero-shot transfer?
Third, can density-only pretraining provide a useful initialization for downstream energy--force learning under a fixed fine-tuning protocol?
Detailed descriptions of the datasets used in these experiments are provided in Supplementary Note~2.

\subsection{Density Reconstruction}
\begin{figure*}
    \centering
    \includegraphics[width=0.9\linewidth]{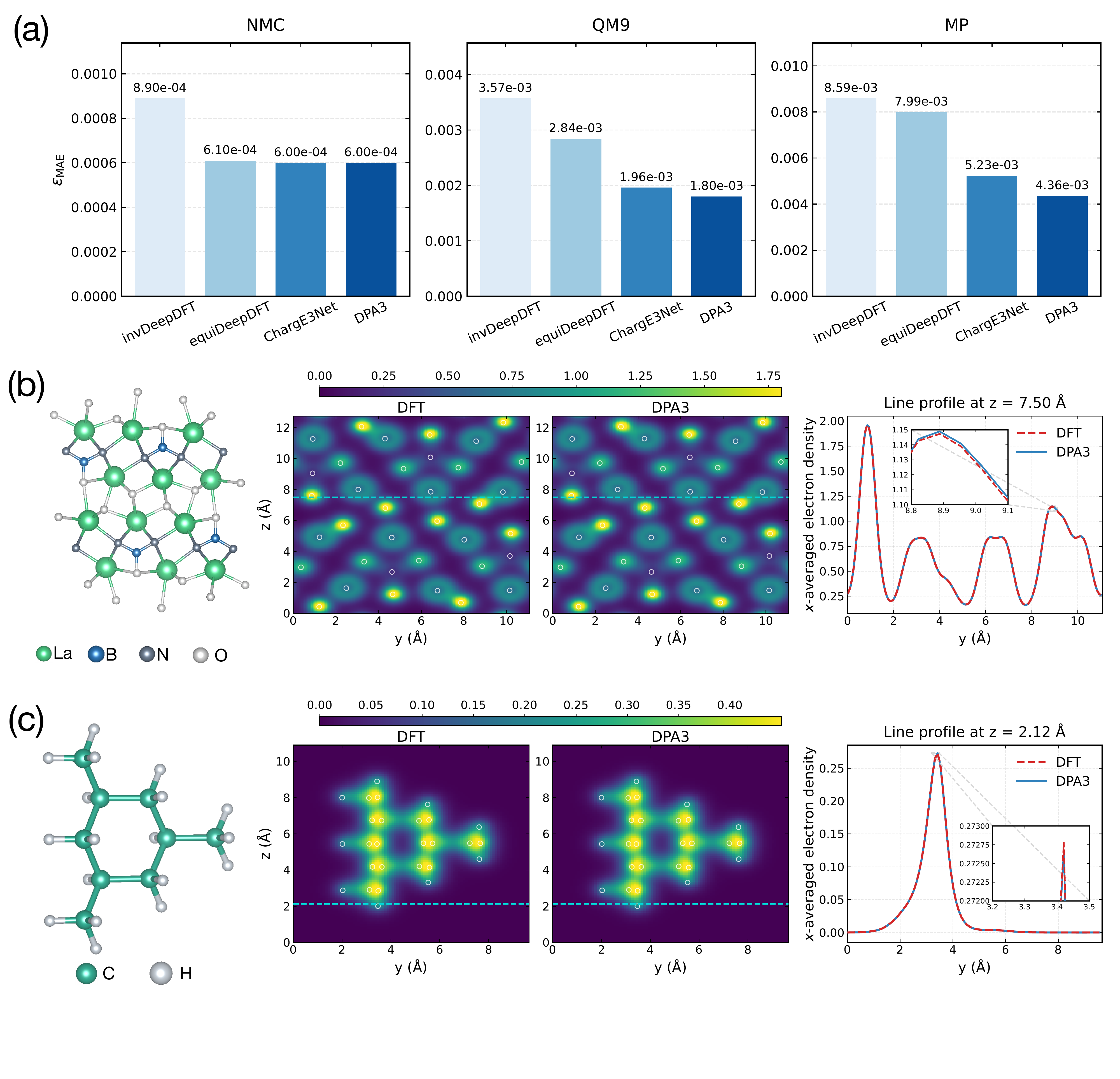}
    \caption{
Electron-density reconstruction with the virtual-atom DPA3 model.
(a) Normalized density MAE, $\epsilon_{\mathrm{MAE}}$, for DPA3 and three reference models on NMC, QM9, and MP.
(b,c) Representative periodic La$_{12}$B$_4$N$_8$O$_{12}$ and molecular C$_9$H$_{18}$ reconstructions.
Each row shows the structure, DFT and DPA3 density maps averaged along $x$, and a line profile along $y$ at the $z$ position marked by the dashed line; circles mark projected atoms and insets enlarge local deviations.}
    \label{fig:figure2}
\end{figure*}

Before examining transfer to interatomic potentials, we first establish that
the virtual-atom formulation can represent spatially resolved electron density
accurately. This provides the basis for using density as a supervision signal
for the shared encoder.
Following Ref.~\citep{koker2024higher}, we adopt the normalized density MAE,
\begin{equation}
\epsilon_{\mathrm{MAE}}
=
\frac{
\sum_{\mathbf{r}_g \in \mathcal{Q}}
\left|
\hat{\rho}(\mathbf{r}_g)-\rho(\mathbf{r}_g)
\right|
}{
\sum_{\mathbf{r}_g \in \mathcal{Q}}
\left|
\rho(\mathbf{r}_g)
\right|
},
\label{eq:reseq1}
\end{equation}
to quantify the reconstruction error. 
Fig.~\ref{fig:figure2}(a) compares DPA3 with representative density-prediction models,
including invDeepDFT~\citep{jorgensen2020deepdft}, equiDeepDFT~\citep{jorgensen2022equivariant}, 
and ChargE3Net~\citep{koker2024higher}, on three benchmarks previously used for charge-density evaluation: NMC, QM9, and MP. 
The NMC dataset consists of Vienna Ab initio Simulation Package (VASP)-computed electron densities for nickel--manganese--cobalt
oxide cathode materials with varying lithium contents, providing a benchmark for crystalline battery materials. 
QM9 contains VASP electron densities for small organic molecules, enabling evaluation on finite molecular systems. 
The MP benchmark is substantially broader, comprising electron densities for a large and chemically diverse collection of inorganic bulk materials from the Materials Project database.

DPA3 achieves competitive or superior reconstruction accuracy across all three datasets.
  On NMC, DPA3 reaches an $\epsilon_{\mathrm{MAE}}$ of $6.00\times10^{-4}$, matching ChargE3Net and outperforming invDeepDFT and equiDeepDFT.
  On QM9, DPA3 obtains an error of $1.80\times10^{-3}$, below the $1.96\times10^{-3}$ reported for ChargE3Net in the comparison shown in Fig.~\ref{fig:figure2}(a).
  Recent specialized models report lower errors on the same split: SCDP reports $1.78\times10^{-3}$~\citep{fu2024recipe} and ELECTRA reports $1.77\times10^{-3}$~\citep{elsborg2025electra}. BOA reports still lower errors on the same QM9--VASP benchmark and the split of Fu et al., reaching $1.381\times10^{-3}$ and $1.339\times10^{-3}$ for its small and large variants, respectively~\citep{klockow2026function}.
  DPA3 should therefore be viewed as competitive, rather than state of the art, on QM9.
  The largest improvement appears on MP, where DPA3 obtains an $\epsilon_{\mathrm{MAE}}$ of $4.36\times10^{-3}$, 
  lower than the $5.23\times10^{-3}$ achieved by ChargE3Net, corresponding to a $16.6\%$ relative
  reduction.
  Because MP spans a broad range of inorganic chemistries and local atomic environments, this result indicates that the virtual-atom DPA3 representation remains effective on a large and chemically diverse density dataset.
  These benchmark results establish that the virtual-atom formulation is competitive with specialized density-prediction architectures across molecular and periodic systems.
  The shared DPA3 encoder therefore provides a sufficiently expressive basis for the subsequent density-pretraining and EF+density experiments.
Detailed training hyperparameters for the density models are provided in Supplementary Table~2.

We further examine the spatial fidelity of the predicted density using 
representative periodic and molecular examples in Fig.~\ref{fig:figure2}(b,c).
For the periodic case, we use the DPA3 model trained on MP and visualize a La$_{12}$B$_4$N$_8$O$_{12}$ structure from the test set.
For the molecular case, we use the DPA3 model trained on QM9 and visualize a C$_9$H$_{18}$ molecule from the test set.
The corresponding normalized density MAEs are $3.64\times10^{-3}$ and $8.81\times10^{-4}$, respectively.
In each example, the DFT reference and DPA3 prediction are shown as two-dimensional density maps averaged along the $x$ direction, 
together with a one-dimensional line profile extracted along the dashed path.

For La$_{12}$B$_4$N$_8$O$_{12}$, DPA3 reproduces the periodic density pattern, including high-density 
regions near atomic sites and smoother variations in interstitial regions.
For C$_9$H$_{18}$, the model captures the localized density peaks around 
the molecular framework as well as the rapid density decay away from the atoms.
The line profiles show that DPA3 accurately follows both the positions 
and magnitudes of the dominant density peaks, with the enlarged insets 
revealing only small local deviations from the DFT reference.
Together, the benchmark comparisons and spatial 
reconstructions demonstrate that the virtual-atom formulation
can accurately reconstruct electron density across molecular and periodic systems.

\subsection{Density-Supervised Multi-Task Learning}
\begin{figure*}
    \centering
    \includegraphics[width=0.9\linewidth]{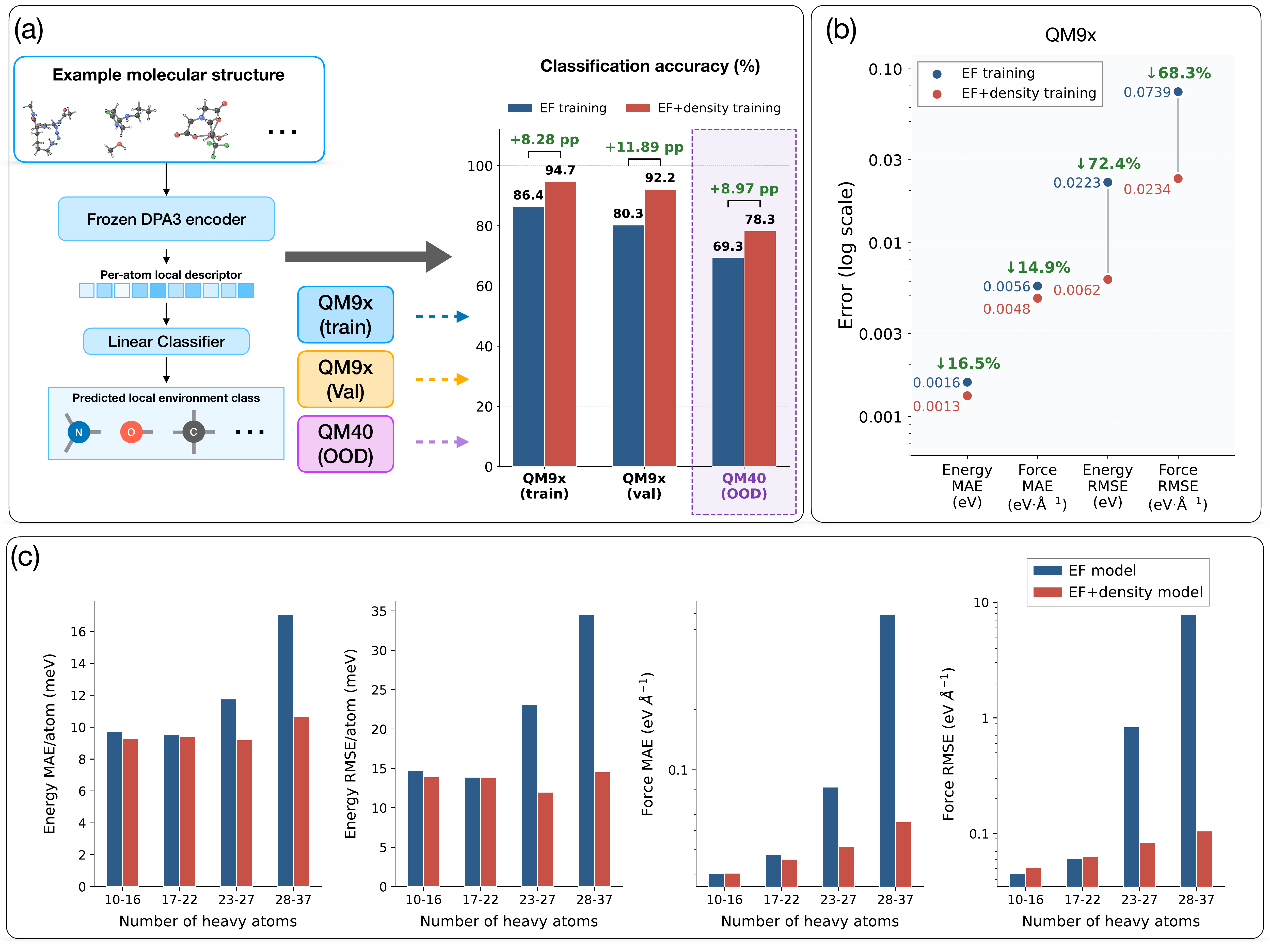}
    \caption{
Effects of multi-task EF+density training on molecular representations and transfer.
(a) Linear-probe accuracy of frozen EF and EF+density encoders on the QM9x training and validation splits and the ORCA-relabeled QM40 subset.
(b) In-domain QM9x energy and force MAEs and RMSEs; percentage reductions use unrounded values, while displayed errors are rounded.
(c) Zero-shot errors on the element-matched, ORCA-relabeled QM40 subset, grouped into the nonoverlapping heavy-atom ranges $10$--$16$, $17$--$22$, $23$--$27$, and $28$--$37$.
Blue and red denote the EF and EF+density models, respectively.
}
    \label{fig:figure3}
\end{figure*}

Having established that the virtual-atom model can reconstruct electron
density, we next ask whether retaining density supervision during energy--force
training improves the shared representation and the resulting potential. We
consider representation quality, in-domain accuracy, and transfer beyond the
molecular-size range used for training.

We compare conventional energy--force (EF) training with multi-task EF+density
training in both periodic and molecular settings.
In the periodic setting, energy and force losses are evaluated on
configurations from the Materials Project Trajectory (MPtrj)
dataset~\citep{deng2023chgnet,deng2023mptrj}, while the density loss is
evaluated on configurations from the MP electron-density dataset. Both
sources derive from Materials Project calculations, so they are not
guaranteed to be configuration-disjoint: MP density data are based primarily
on relaxed or static structures, whereas MPtrj also contains relaxation
trajectory frames. The corresponding MP/MPtrj in-domain results and
cross-domain evaluations are reported in Supplementary Fig.~1 and
Supplementary Table~1, respectively.

In the molecular setting, energy and force labels are obtained from QM9x,
while density labels are obtained from the QM9 electron-density dataset.
QM9x recomputes the configurations in QM9 at the $\omega$B97X/6-31G(d) level and provides
energies and nonzero atomic forces~\citep{schreiner2022qm9x,schreiner2022transition1x}, while
the QM9 density data use the QM9 molecular geometries. The two supervision
sources therefore have substantial molecular-identity and geometry overlap,
although preprocessing and split assignments need not be identical. This is
the intended multi-task setting: density provides an additional
electronic-structure target, rather than an energy--force label for the same
sample. During EF+density training, each mixed batch contains 256 QM9x energy--force 
structures and 256 QM9 density structures; the EF-only baseline contains 256 QM9x structures per step.
Because both models are trained for $2\times10^6$ steps, they see the same total 
number of energy--force sample presentations, and the larger EF+density batch reflects only the additional density supervision.
Figure~\ref{fig:figure3} focuses on this molecular setting, for which we
perform representation probing, in-domain evaluation, and zero-shot
molecular transfer.

To assess the learned representation, we freeze the EF and EF+density
encoders, extract per-atom descriptors, and train a linear classifier to
predict each atom's local-environment class (Fig.~\ref{fig:figure3}(a)).
Because the encoders are fixed, the result measures how accessible this
information is from the learned descriptors.

Density supervision increases the classification accuracy from $86.4\%$ to
$94.7\%$ on the QM9x training split and from $80.3\%$ to $92.2\%$ on the
validation split. The improvement also persists outside the training distribution:
on the ORCA-relabeled QM40 subset~\citep{madushanka2024qm40}, the
accuracy increases from $69.3\%$ to $78.3\%$. These results indicate that
density supervision makes local-environment information more accessible to a
linear readout, including on molecular environments outside the training
domain.

As shown in
Fig.~\ref{fig:figure3}(b), adding QM9 density supervision to QM9x
energy--force training reduces the energy MAE, force MAE, energy
root-mean-square error (RMSE), and force RMSE by $16.5\%$, $14.9\%$, $72.4\%$,
and $68.3\%$, respectively. The representation-level improvement is therefore
accompanied by lower energy and force errors on QM9x.
The larger reductions in RMSE than in MAE suggest that the improvement is
particularly pronounced for configurations with larger prediction errors.
In the periodic MP/MPtrj setting, density supervision produces
smaller reductions across the same four metrics in the reported run, as reported
in Supplementary Fig.~1. The smaller magnitude of these improvements may
reflect the heterogeneous electronic-structure settings of the MP density
data, including Hubbard-$U$ corrections for many systems, as well as the
absence of configuration-wise pairing between the MP density and MPtrj
energy--force labels.

We further evaluate the two periodic checkpoints without fine-tuning on
subsets of MatPES, OC20, ODAC23, OMol25, and QCML recomputed under periodic
boundary conditions (PBC). These external evaluation configurations are not
used to train either checkpoint and therefore probe transfer beyond the
source MP/MPtrj structures. The EF+density model achieves a lower force MAE on all five external datasets,
although the change is marginal on ODAC23, whereas the effect on energy MAE is 
dataset dependent: energy errors decrease on OMol25 and QCML but increase 
on MatPES, OC20, and ODAC23 (Supplementary Table~1).
Within these zero-shot periodic evaluations, the benefit of density supervision 
is therefore more robust for force prediction than for energy prediction.

Finally, we examine whether the molecular representations transfer beyond
the molecular-size range encountered during training. QM9x contains
molecules with at most nine heavy atoms. We therefore construct an
out-of-distribution evaluation set by randomly sampling molecules from QM40
whose heavy-atom counts are greater than nine. To remove extrapolation to
unseen elements as one confounding factor, we retain only molecules composed
of the same elements represented in QM9 and QM9x, namely H, C, N, O, and F.
Because every QM9x molecule contains at most nine heavy atoms, these larger
QM40 molecules are outside the molecular-size range of both training sources.
The resulting set is therefore an element-matched out-of-distribution test
that includes changes in molecular size, composition, and bonding environments.

Because the original QM40 labels were generated using electronic-structure
settings different from those used for QM9x, we recompute the energies and
atomic forces of the selected molecules using ORCA at the same nominal level
of theory as the QM9x energy--force labels. We refer to this evaluation set
as the ORCA-relabeled QM40 subset. None of these configurations or their
relabeled energy and force values are used for model training, validation,
checkpoint selection, or fine-tuning.

We directly evaluate the molecular EF and EF+density checkpoints on the
ORCA-relabeled QM40 subset and group the prediction errors according to the
number of heavy atoms, as shown in Fig.~\ref{fig:figure3}(c). The
corresponding numerical values for the four nonoverlapping size intervals
are reported in Supplementary Table~5.

For molecules in the two smaller size groups, the models have similar energy errors and force MAEs, 
while the EF+density force RMSE is slightly higher. Their behavior diverges as molecular size
increases. In the largest group, containing $28$--$37$ heavy atoms, density
supervision reduces the force MAE from $5.97\times10^{-1}$ to
$5.50\times10^{-2}$~eV~\AA$^{-1}$ and the force RMSE from $7.90$ to
$1.05\times10^{-1}$~eV~\AA$^{-1}$, corresponding to reductions of $90.8\%$
and $98.7\%$, respectively. The energy MAE and RMSE per atom decrease by
$37.3\%$ and $57.8\%$ in the same size group. The particularly large force
RMSE of the EF model suggests that high-error cases contribute substantially
to this metric; the simultaneous reduction in force MAE indicates that the
improvement is not confined to RMSE. The
advantage of density supervision is therefore most pronounced for larger
molecules and for atomic-force prediction.

Together, the linear-probe, in-domain, and zero-shot results show that density
supervision makes local-environment information more accessible, improves
energy and force prediction within the training domain, and is particularly
helpful for force prediction beyond that domain. This empirical pattern is consistent with the
relationship between electron density and atomic forces discussed in
Eq.~\ref{eq:force_density_relation}: reconstructing the density may encourage
the shared encoder to retain spatially resolved electronic structure
information that remains relevant as local environments and molecular sizes
move beyond the training distribution. Although these experiments do not
establish the microscopic mechanism responsible for the improvement, they
provide evidence that electron density can serve as a transferable supervision
signal for molecular representation learning.

Figure~\ref{fig:figure3} establishes the central result: retaining
density supervision during energy--force training improves the shared
representation and transfer performance. We then ask whether an encoder
trained only on density can provide a useful initialization for downstream
energy--force learning.

\subsection{Transfer from Density Pretraining}
\begin{figure*}
      \centering
      \includegraphics[width=0.9\linewidth]{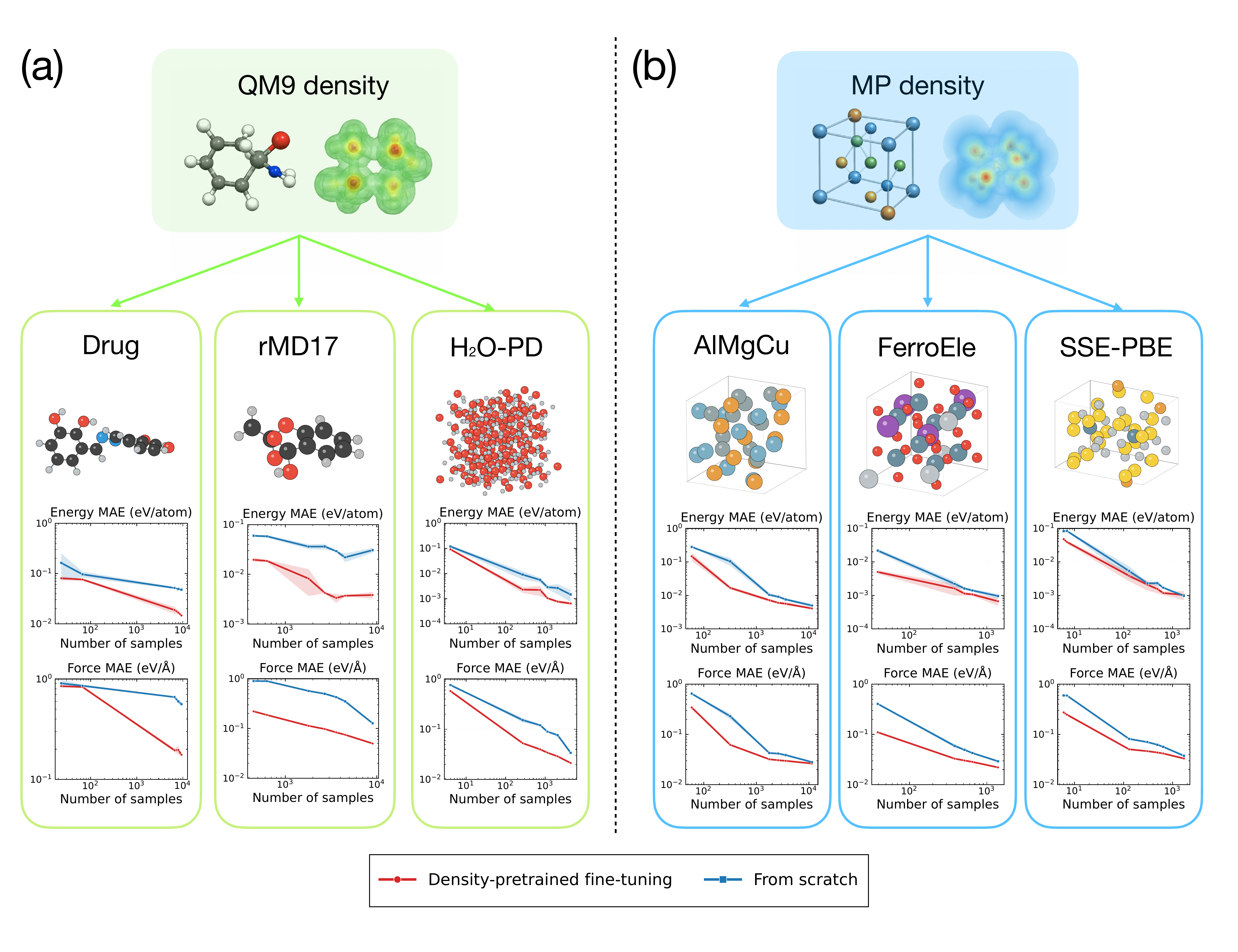}
      \caption{
Transfer from density pretraining to energy--force learning.
(a) Molecular tasks (drug, rMD17, and H$_2$O-PD) initialized from a QM9-density checkpoint.
(b) Solid-state tasks (AlMgCu, FerroEle, and SSE-PBE) initialized from an MP-density checkpoint.
Each column shows a representative structure and the validation energy MAE per atom and force MAE versus the number of downstream training samples.
Red denotes density-pretrained initialization and blue denotes training from scratch under the same fixed-epoch protocol.
Curves show the mean over three independent random seeds, and shaded regions indicate one standard deviation.
}
      \label{fig:figure4}
  \end{figure*}

Recent work on universal and pretrained atomistic models has shown that
representations learned from large source datasets can provide useful
initializations for downstream fine-tuning~\citep{chen2022universal,deng2023chgnet,yang2024mattersim,zhang2024dpa}.
Here, density labels shape the encoder during pretraining but are absent from
downstream energy--force optimization. We therefore test whether density-only
pretraining provides a useful initialization under a fixed fine-tuning protocol;
the experiment is intended to assess transfer, not to determine energy or force
accuracy from density reconstruction alone.

To evaluate this idea, we construct density-pretrained checkpoints for molecular and solid-state domains.
For molecular downstream tasks, including drug~\citep{zhang2024dpa}, 
revised MD17 (rMD17)~\citep{christensen2020role}, and H$_2$O-PD~\citep{zhang2021phase},
we fine-tune from a checkpoint pretrained on QM9 electron densities.
For solid-state downstream tasks, including AlMgCu~\citep{jiang2021accurate}, 
FerroEle~\citep{wu2023universal}, and SSE-PBE~\citep{huang2021deep},
we fine-tune from a checkpoint pretrained on MP electron
densities.
Each pretrained model is optimized using only downstream energy and 
force labels and is compared with a model trained from scratch under the same data budget.
For each downstream dataset, the density-pretrained model and the 
from-scratch baseline are trained for the same number of effective epochs.
Detailed fine-tuning settings, including the pretrained checkpoint 
and the corresponding number of training epochs, are provided in Supplementary Table~4.

Figure~\ref{fig:figure4} shows the validation energy and force MAEs as
functions of the number of labeled downstream training samples. Density
pretraining generally lowers the validation errors, particularly at smaller
data budgets, although the size of the benefit depends on the observable and
the downstream task. For the molecular tasks, the improvement is most
consistent for forces, while the energy curves show more task-dependent
behavior. On the solid-state tasks, the pretrained models have lower errors
at small data budgets and the curves tend to converge as more downstream data
become available. These results show that density-only checkpoints can be
adapted to energy--force learning under the evaluated protocol.

The stronger and more consistent force improvements are consistent with the
fact that density reconstruction exposes spatial variations in the local
electronic environment, whereas energy is a global scalar. This interpretation
is compatible with the relationship in Eq.~\ref{eq:force_density_relation},
but that relationship does not determine the observed error reductions or
their microscopic origin. Taken together, the six tasks show that density
reconstruction can provide a reusable initialization for energy--force
learning under the protocol examined here, supporting the interpretation of
the multi-task results in Fig.~\ref{fig:figure3}.

\section{Discussion and Conclusion}

In this work, we introduced an asymmetric virtual-atom formulation that makes
electron-density learning compatible with the DPA3 encoder used for interatomic
potentials. Virtual atoms provide density-query centers without changing the
real-atom interaction graph, allowing density and energy--force tasks to share
encoder parameters while retaining separate readouts.

The experiments show two related uses of density supervision. When density and
energy--force labels are used jointly, density improves the shared
representation and downstream prediction. When density is used for pretraining
only, the resulting checkpoint can be adapted to molecular and solid-state
energy--force tasks under the fixed fine-tuning protocol. The improvements are
more consistent for forces in the periodic-transfer and molecular zero-shot
experiments, although both energies and forces improve in several in-domain and
fine-tuning settings.

This force-dominant pattern is physically plausible: density describes spatial
variation in the electronic environment, whereas energy is a global scalar.
The QM40 results provide the clearest example, with the benefit of density
supervision increasing as molecular size moves beyond the QM9x range. Together
with the linear-probe results, they suggest that density-supervised
representations retain useful local-environment information outside the source
distribution. These observations are consistent with, but do not establish,
a microscopic causal mechanism.

Several limitations remain. Density reconstruction requires evaluating virtual
atoms at spatial query points, which adds inference cost when the density head
is used, although this cost disappears after the head is discarded in the
pretraining setting. The EF and EF+density experiments match the number of
energy--force sample presentations but not total computational cost, since the
latter also processes density structures and uses more GPUs. The current
framework treats density as a supervised target rather than coupling it to
energy through a self-consistent or variational relation. More efficient
density sampling, improved loss balancing, and extensions to charge transfer,
polarization, and long-range interactions are promising directions for future
work.

Overall, the virtual-atom formulation turns electron density from an isolated
field-reconstruction target into a supervision signal for interatomic-potential
representations. Its benefits are observed in density reconstruction,
fine-tuning, joint energy--force learning, representation probing, and
zero-shot molecular transfer.

\section{Data Availability}
The data supporting the findings of this study are available on AIS Square in the
\textit{DPA3\_density\_paper} dataset (ID 440)~\citep{liu2026dpa3_density_dataset}.

\section{Code Availability}
The grid-based charge-density implementation developed for this work, including
training, inference, evaluation, and DPA3 example configurations, is available in
DeePMD-kit pull request \#5999~\citep{liu2026deepmd_density}.

\section{Acknowledgments}
We thank Nan Sheng for insightful discussions.
Computational resources were provided by the AI for Science Institute.
This work was supported by the National Key R\&D Program of China (Grant No.~2022YFA1004300)
and the National Natural Science Foundation of China (Grant Nos.~12525113 and 12561160120).

\bibliography{sn-bibliography}
\end{document}


\tableofcontents

\section{Periodic Multi-Task Learning}
\label{sec:periodic_joint_training}

In addition to the molecular QM9/QM9x experiments presented in the main
text, we evaluate multi-task EF+density training in a periodic materials
setting. Energy and force supervision is provided by configurations from the
Materials Project Trajectory (MPtrj)
dataset~\citep{deng2023chgnet,deng2023mptrj}, whereas electron-density
supervision is obtained from the MP density dataset. Both sources derive from
Materials Project calculations and are therefore not
guaranteed to be configuration-disjoint: MP density data are based primarily
on relaxed or static structures, whereas MPtrj also contains
relaxation-trajectory frames. The labels are not paired frame by frame, but
the two datasets can share material identities or closely related structures.

We compare conventional energy--force (EF) training on MPtrj with multi-task
EF+density training, in which the model additionally receives
density supervision from the MP density dataset. The two models use the same
DPA3 architecture and energy--force loss settings. During EF+density training,
MPtrj energy--force batches and MP density batches are used jointly, with each
loss evaluated only on samples carrying the corresponding labels.

\begin{figure*}[htbp]
    \centering
    \includegraphics[width=0.7\linewidth]{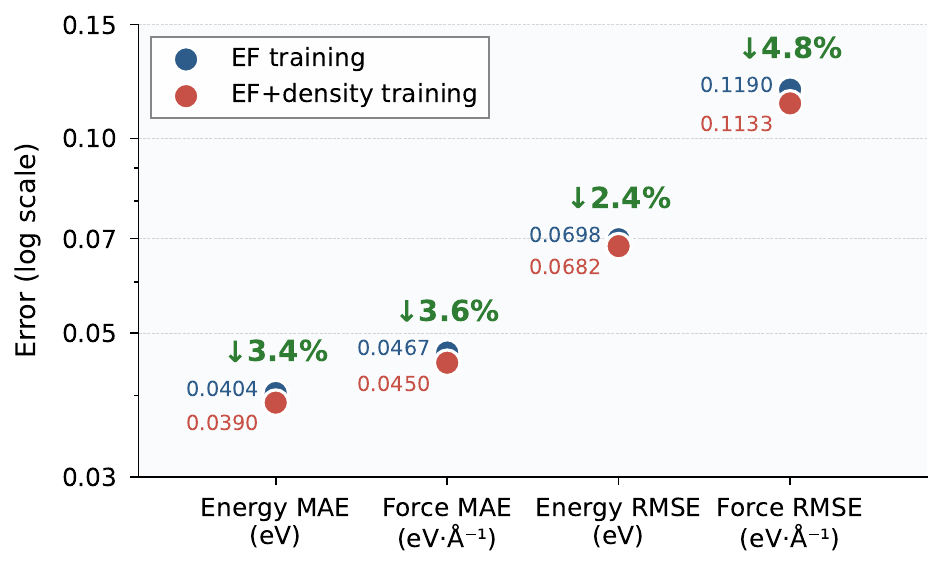}
    \caption{In-domain MPtrj energy and force MAEs and RMSEs for EF and multi-task EF+density training.
    Both models use MPtrj energy--force labels; EF+density additionally uses MP density labels.
    Percentage reductions use unrounded results, displayed errors are rounded, and the vertical axis is logarithmic.
    Blue and red denote EF and EF+density, respectively.}
    \label{fig:mptrj_in_domain}
\end{figure*}

As shown in Supplementary Fig.~\ref{fig:mptrj_in_domain}, density supervision
reduces all four in-domain metrics in the reported run. The energy MAE decreases from
$0.0404$ to $0.0390$~eV, while the force MAE decreases from $0.0467$ to
$0.0450$~eV~\AA$^{-1}$. The energy RMSE decreases from $0.0698$ to
$0.0682$~eV, and the force RMSE decreases from $0.1190$ to
$0.1133$~eV~\AA$^{-1}$. Based on the unrounded numerical values, these
changes correspond to reductions of approximately $3.4\%$, $3.6\%$,
$2.4\%$, and $4.8\%$, respectively. The differences are modest and would
require repeated runs to distinguish them from seed-to-seed variation.

The comparatively modest improvements in the MP/MPtrj setting may reflect two
characteristics of the training data. First, the MP density labels are not
generated under a fully uniform electronic-structure protocol: a substantial
fraction of the calculations employ Hubbard-$U$ corrections, and the detailed
settings vary across chemical systems. Second, the MP density and MPtrj
datasets do not provide configuration-wise paired density and energy--force
labels. The two objectives are instead evaluated on different structures drawn
from related but nonidentical distributions. Consequently, density supervision
constrains the shared representation at the dataset level rather than providing
a density label for each MPtrj trajectory frame. These factors may weaken the
alignment between the density and energy--force objectives and help explain why
the improvements are smaller than those observed in the more closely matched
QM9/QM9x setting.

To determine whether this improvement transfers beyond the MPtrj
distribution, we further evaluate the two checkpoints on external datasets
covering several distinct chemical domains. The evaluation sets are derived
from MatPES~\citep{kaplan2025foundational}, OC20~\citep{chanussot2021open},
ODAC23~\citep{sriram2025open}, OMol25~\citep{levine2025open}, and
QCML~\citep{ganscha2025qcml}. We use the subsets released with the
SevenNet-Omni study~\citep{kim2026optimizing,kim2026domainbridging}, in which
selected configurations from these datasets were recomputed under a unified
density functional theory (DFT) protocol with periodic boundary conditions
(PBC). Neither model is fine-tuned on these target datasets before evaluation.

\begin{table*}[htbp]
    \centering
    \caption{Zero-shot energy and force errors of the MPtrj-trained EF and
    EF+density models on the PBC-recomputed external datasets. Energy MAEs are
    reported per atom. No target-specific fine-tuning is performed. The force
    reductions are calculated from the unrounded numerical results; the MAEs
    shown in the table are rounded to four decimal places.}
    \label{tab:mptrj_cross_domain}
    \small
    \setlength{\tabcolsep}{7pt}
    \renewcommand{\arraystretch}{1.15}
    \begin{tabular}{lccccc}
        \toprule
        & \multicolumn{2}{c}{Energy MAE/atom (eV)}
        & \multicolumn{2}{c}{Force MAE (eV~\AA$^{-1}$)}
        & Force MAE $\downarrow$ \\
        \cmidrule(lr){2-3}\cmidrule(lr){4-5}
        Dataset & EF & EF+density & EF & EF+density & (\%) \\
        \midrule
        MatPES & 0.0917 & 0.2136 & 0.1765 & 0.1622 & 8.05 \\
        OC20   & 0.0343 & 0.0424 & 0.1981 & 0.1752 & 11.56 \\
        ODAC23 & 0.0274 & 0.0293 & 0.1430 & 0.1425 & 0.33 \\
        OMol25 & 0.0252 & 0.0240 & 0.2062 & 0.1899 & 7.92 \\
        QCML   & 0.1687 & 0.1479 & 0.3356 & 0.3225 & 3.92 \\
        \bottomrule
    \end{tabular}
\end{table*}

Supplementary Table~\ref{tab:mptrj_cross_domain} shows that the EF+density
model consistently reduces the force MAE across all five external datasets,
with the most pronounced improvement on OC20 and only a marginal change on
ODAC23. In contrast, the effect on energy prediction is dataset dependent: the
energy MAE decreases on OMol25 and QCML but increases on MatPES, OC20, and
ODAC23; on MatPES, it more than doubles. These results suggest that the cross-domain benefit of density
supervision is more robust for force prediction than for energy prediction.

The in-domain changes are modest in the reported run, whereas the external
evaluations show lower force MAEs across all five tested domains, with only a
marginal change on ODAC23. Energy transfer remains dataset dependent.

\section{Dataset Descriptions}

\subsection{Electron-Density Datasets}

This section describes the datasets used in the \emph{Electron-Density Reconstruction} section of the main text, where we benchmark the density prediction accuracy of the proposed virtual-atom formulation.

\textbf{NMC.}
The NMC dataset consists of lithium-ion battery cathode materials based on nickel--manganese--cobalt transition-metal oxides~\citep{jorgensen2022nmc}.
It provides DFT-computed electron densities together with the corresponding atomic structures and energies.
Following previous density-prediction benchmarks~\citep{jorgensen2022equivariant,koker2024higher}, we use the original training, validation, and test splits for evaluating density reconstruction performance.

\textbf{QM9.}
The QM9 dataset contains equilibrium geometries of small organic molecules composed of H, C, N, O, and F atoms~\citep{ramakrishnan2014quantum}.
The DFT-computed electron densities are represented on uniform real-space grids.
We use the processed QM9 charge-density dataset from Ref.~\citep{jorgensen2022qm9}, which provides the training, validation, and test splits used in our benchmark.

\textbf{MP.}
The MP electron-density dataset is constructed from crystalline materials in the Materials Project database~\citep{shen2022representation}.
Compared with NMC and QM9, this dataset covers a broader range of inorganic compositions and periodic crystal structures.
We use the publicly available training, validation, and test splits provided by Ref.~\citep{koker2024higher}.

\subsection{Energy and Force Datasets}

This section describes the downstream energy and force datasets used in the
\emph{Transfer from Density Pretraining}
and
\emph{Density-Supervised Multi-Task Learning}
sections of the main text.

The datasets used in Fig.~4 include AlMgCu, FerroEle, SSE-PBE, drug, H$_2$O-PD, and revised MD17 (rMD17).
AlMgCu, FerroEle, SSE-PBE, drug, and H$_2$O-PD are obtained from the OpenLAM-TrainingSet-v1 dataset hosted on AIS Square~\citep{openlam2025training}.
These datasets cover both solid-state and molecular systems and are used to evaluate density-pretrained initialization under a fixed fine-tuning protocol across different downstream energy--force learning scenarios.
The rMD17 dataset~\citep{christensen2020rmd17} provides DFT energies and forces for molecular dynamics trajectories of small organic molecules.
For all downstream fine-tuning datasets, we randomly split the available configurations into training and validation sets with a 9:1 ratio.

The multi-task EF+density experiments in Fig.~3 use MPtrj and QM9x as the energy--force sources and the MP and QM9 electron-density datasets described above as the density sources.
MPtrj is derived from Materials Project relaxation trajectories~\citep{deng2023chgnet,deng2023mptrj}, while QM9x recomputes the molecular geometries underlying QM9 and provides both energies and nonzero atomic forces~\citep{schreiner2022transition1x}.
Accordingly, the QM9/QM9x and MP/MPtrj source pairs can share molecular or
material identities and closely related structures across tasks. We interpret
the in-domain comparisons as evaluating the benefit of an additional density
target within a shared structural domain; the zero-shot evaluations use
held-out configurations from external datasets.
The molecular out-of-distribution evaluation in Fig.~3 uses a randomly
sampled subset of QM40~\citep{madushanka2024qm40}. We retain only molecules
containing more than nine heavy atoms and composed of the elements present in
QM9/QM9x (H, C, N, O, and F). Their energies and atomic forces are recomputed
with ORCA at the same nominal level of theory as the QM9x energy--force
labels.

\section{Training Configurations}

The hyperparameters and system configurations are summarized in
Tables~\ref{tab:density_hyperparams} and \ref{tab:single_multi_hyperparams}.
For density reconstruction, the architecture, optimizer, learning-rate
schedule, training length, and density-loss prefactor are shared across NMC,
QM9, and MP; only the number of GPUs is adjusted. Each density-only model is
optimized for $2\times10^6$ steps with a global batch size of 256, corresponding
to $5.12\times10^8$ density-structure presentations.

The EF and EF+density models use the same backbone, optimizer, learning-rate
schedule, number of steps, and energy/force loss prefactors. Each step contains
256 energy--force structures; EF+density adds 256 density structures, giving a
total batch size of 512 without increasing the energy--force label budget.
After $2\times10^6$ steps, both models have seen $5.12\times10^8$
energy--force structures. Batch sizes below are global totals across all GPUs.

\begin{table*}[htbp]
  \centering
  \begin{threeparttable}
  \caption{Hyperparameters for density training on NMC, QM9, and MP datasets.}
  \label{tab:density_hyperparams}
  \setlength{\tabcolsep}{14pt}  
  \renewcommand{\arraystretch}{1.2} 
  \setlength{\arrayrulewidth}{0.6pt}  
  \begin{tabular}{l|ccc}
    \toprule
    \textbf{Hyperparameters} &
    \textbf{NMC} &
    \textbf{QM9} &
    \textbf{MP} \\
    \hline
    Number of update layers $L$ & 12 & 12 & 12 \\
    Maximum graph order $k$ & 2 & 2 & 2 \\
    Dimension of vertex features in $G^{(0)}$ & 128 & 128 & 128 \\
    Dimension of vertex features in $G^{(1)}$ & 64 & 64 & 64 \\
    Dimension of vertex features in $G^{(2)}$ & 32 & 32 & 32 \\
    Cutoff radius in $G^{(1)}$ ($\text{\AA}$) & 6.0 & 6.0 & 6.0 \\
    Cutoff radius in $G^{(2)}$ ($\text{\AA}$) & 4.0 & 4.0 & 4.0 \\
    Batch size & 256 & 256 & 256 \\
    Number of GPUs & 2 & 2 & 4 \\
    Optimizer & AdamW & AdamW & AdamW \\
    Learning rate scheduling & Exp & Exp & Exp \\
    Maximum learning rate & $1\times10^{-3}$ & $1\times10^{-3}$ & $1\times10^{-3}$ \\
    Minimum learning rate & $1\times10^{-5}$ & $1\times10^{-5}$ & $1\times10^{-5}$ \\
    Number of training steps & 2M & 2M & 2M \\
    Total density-structure sample exposures & $5.12\times10^8$ & $5.12\times10^8$ & $5.12\times10^8$ \\
    Activation function threshold & 10.0 & 10.0 & 10.0 \\
    Density loss prefactor & $1 \rightarrow 1$ & $1 \rightarrow 1$ & $1 \rightarrow 1$ \\
    \bottomrule
  \end{tabular}
  \end{threeparttable}
\end{table*}

\begin{table*}[htbp]
  \centering
  \begin{threeparttable}
  \caption{Hyperparameters for EF and multi-task EF+density training on QM9x/QM9 and MPtrj/MP.}
  \label{tab:single_multi_hyperparams}
  \setlength{\tabcolsep}{7pt}
  \renewcommand{\arraystretch}{1.2} 
  \setlength{\arrayrulewidth}{0.6pt}  
  \begin{tabular}{l|cc|cc}
    \toprule
    \textbf{Hyperparameters} &
    \multicolumn{2}{c|}{\textbf{QM9x / QM9}} &
    \multicolumn{2}{c}{\textbf{MPtrj / MP}} \\
    \cline{2-5}
    &
    \textbf{EF} &
    \textbf{EF+density} &
    \textbf{EF} &
    \textbf{EF+density} \\
    \hline
    Number of update layers $L$ & 12 & 12 & 12 & 12 \\
    Maximum graph order $k$ & 2 & 2 & 2 & 2 \\
    Dimension of vertex features in $G^{(0)}$ & 128 & 128 & 128 & 128 \\
    Dimension of vertex features in $G^{(1)}$ & 64 & 64 & 64 & 64 \\
    Dimension of vertex features in $G^{(2)}$ & 32 & 32 & 32 & 32 \\
    Cutoff radius in $G^{(1)}$ ($\text{\AA}$) & 6.0 & 6.0 & 6.0 & 6.0 \\
    Cutoff radius in $G^{(2)}$ ($\text{\AA}$) & 4.0 & 4.0 & 4.0 & 4.0 \\
    Energy--force batch size & 256 & 256 & 256 & 256 \\
    Density batch size & -- & 256 & -- & 256 \\
    Total batch size & 256 & 512 & 256 & 512 \\
    Number of GPUs & 2 & 4 & 4 & 8 \\
    Optimizer & AdamW & AdamW & AdamW & AdamW \\
    Learning rate scheduling & Exp & Exp & Exp & Exp \\
    Maximum learning rate & $1\times10^{-3}$ & $1\times10^{-3}$ & $1\times10^{-3}$ & $1\times10^{-3}$ \\
    Minimum learning rate & $1\times10^{-5}$ & $1\times10^{-5}$ & $1\times10^{-5}$ & $1\times10^{-5}$ \\
    Number of training steps & 2M & 2M & 2M & 2M \\
    Total energy--force sample exposures & $5.12\times10^8$ & $5.12\times10^8$ & $5.12\times10^8$ & $5.12\times10^8$ \\
    Total density sample exposures & -- & $5.12\times10^8$ & -- & $5.12\times10^8$ \\
    Activation function threshold & 10.0 & 10.0 & 10.0 & 10.0 \\
    Energy loss prefactor & $0.2 \rightarrow 20$ & $0.2 \rightarrow 20$ & $0.2 \rightarrow 20$ & $0.2 \rightarrow 20$ \\
    Force loss prefactor & $100 \rightarrow 60$ & $100 \rightarrow 60$ & $100 \rightarrow 60$ & $100 \rightarrow 60$ \\
    Density loss prefactor & -- & $1 \rightarrow 1$ & -- & $1 \rightarrow 1$ \\
    \bottomrule
  \end{tabular}
  \begin{tablenotes}
    \footnotesize
    \item EF denotes conventional energy--force training; EF+density denotes multi-task training with an additional density batch. The two settings match the energy--force sample budget but not the total computational cost, because EF+density training additionally processes density structures.
  \end{tablenotes}
  \end{threeparttable}
\end{table*}

\section{Density Pretraining and Fine-Tuning}

For downstream fine-tuning experiments, we initialize the models from density-pretrained checkpoints and adapt them to target datasets with limited training samples. 
Specifically, the checkpoint pretrained on the MP density dataset is used for AlMgCu, FerroEle, and SSE-PBE, while the checkpoint pretrained on the QM9 density dataset is used for Drug, H$_2$O-PD, and rMD17. 
The number of effective fine-tuning epochs and the corresponding pretrained checkpoints for each downstream dataset are summarized in Table~\ref{tab:finetune_epochs}. 
For each dataset, the training samples are randomly selected from the corresponding training set, and all fine-tuning experiments use a batch size of 128. 
When varying the number of training samples, the number of optimization steps is adjusted according to the dataset size so that different sample-size settings correspond to the same number of effective epochs.
For a downstream training subset containing $N_{\mathrm{train}}$ configurations, batch size 128, and the dataset-specific epoch count $E_d$ listed in Table~\ref{tab:finetune_epochs}, the number of optimizer updates is
$E_d\lceil N_{\mathrm{train}}/128\rceil$.
Each learning-curve point is trained independently with three random seeds; curves report the mean and shaded regions show one standard deviation.
All models are evaluated using the final checkpoint after the prescribed number of epochs.
The validation split is used only for evaluation and is not used to select checkpoints or stop training.

\begin{table}[htbp]
  \centering
  \begin{threeparttable}
  \caption{Pretrained checkpoints and effective fine-tuning epochs for downstream datasets.}
  \label{tab:finetune_epochs}
  \setlength{\tabcolsep}{12pt}
  \renewcommand{\arraystretch}{1.15}
  \setlength{\arrayrulewidth}{0.6pt}
  \begin{tabular}{lcc}
    \toprule
    \textbf{Downstream dataset} & 
    \textbf{Pretrained checkpoint} & 
    \textbf{Fine-tuning epochs} \\
    \hline
    AlMgCu   & MP density  & 40  \\
    FerroEle & MP density  & 164 \\
    SSE-PBE  & MP density  & 44  \\
    Drug     & QM9 density & 2   \\
    H$_2$O-PD & QM9 density & 13  \\
    rMD17    & QM9 density & 9   \\
    \bottomrule
  \end{tabular}
  \end{threeparttable}
\end{table}

\section{QM40 Out-of-Distribution Evaluation}
\label{sec:qm40_ood}

Table~\ref{tab:qm40_ood} reports the numerical values underlying Fig.~3(c) of
the main text, that is, the zero-shot energy and force errors of the
energy--force (EF) and multi-task EF+density models on
the ORCA-relabeled QM40 subset, grouped into nonoverlapping heavy-atom intervals. Both checkpoints are trained on
QM9x (and QM9 density) only and are evaluated on QM40 without any
fine-tuning.

\begin{table}[htbp]
  \centering
  \begin{threeparttable}
  \caption{Zero-shot energy and force errors of the EF and EF+density models on the ORCA-relabeled QM40 subset, grouped into nonoverlapping heavy-atom intervals. Energy errors are reported per atom in meV, and force errors in eV~\AA$^{-1}$.}
  \label{tab:qm40_ood}
  \setlength{\tabcolsep}{8pt}
  \renewcommand{\arraystretch}{1.15}
  \setlength{\arrayrulewidth}{0.6pt}
  \begin{tabular}{cc|cc|cc}
    \toprule
    \textbf{Heavy atoms} & \textbf{Model} &
    \textbf{E MAE/atom} & \textbf{E RMSE/atom} &
    \textbf{F MAE} & \textbf{F RMSE} \\
    & & (meV) & (meV) & (eV~\AA$^{-1}$) & (eV~\AA$^{-1}$) \\
    \hline
    \multirow{2}{*}{$10$--$16$}
      & EF         & 9.725 & 14.770 & $3.04\times10^{-2}$ & $4.52\times10^{-2}$ \\
      & EF+density & 9.279 & 13.926 & $3.06\times10^{-2}$ & $5.10\times10^{-2}$ \\
    \hline
    \multirow{2}{*}{$17$--$22$}
      & EF         & 9.546 & 13.888 & $3.80\times10^{-2}$ & $6.10\times10^{-2}$ \\
      & EF+density & 9.400 & 13.782 & $3.59\times10^{-2}$ & $6.34\times10^{-2}$ \\
    \hline
    \multirow{2}{*}{$23$--$27$}
      & EF         & 11.770 & 23.139 & $8.20\times10^{-2}$ & $8.37\times10^{-1}$ \\
      & EF+density &  9.198 & 11.977 & $4.16\times10^{-2}$ & $8.38\times10^{-2}$ \\
    \hline
    \multirow{2}{*}{$28$--$37$}
      & EF         & 17.054 & 34.523 & $5.97\times10^{-1}$ & $7.90$ \\
      & EF+density & 10.688 & 14.559 & $5.50\times10^{-2}$ & $1.05\times10^{-1}$ \\
    \bottomrule
  \end{tabular}
  \end{threeparttable}
\end{table}

\renewcommand{\refname}{Supplementary References}
\bibliography{suppl/supplement}